\def\FullBox{\protect\rule{1.5mm}{1.5mm}}
\def\tipo{2}

\ifnum \tipo = 2
 \documentstyle[prl,twocolumn,aps,psfig]{revtex}
 
 \def\figsiz1{8cm}
 \def \frontmatter{\twocolumn[\hsize\textwidth\columnwidth\hsize\csname@twocolumnfalse\endcsname}

\else
 \documentstyle[preprint,aps,prl,psfig]{revtex}
 
 \def\figsiz1{12cm}
 \def\frontmatter{}
 
\fi

\begin{document}
\draft
\frontmatter
\title{Nonlinearity and Multifractality of Climate Change in the Past
420,000 Years}
\author{Yosef~Ashkenazy$^1$, Don~R.~Baker$^{2,3}$, Hezi~Gildor$^4$,
Shlomo~Havlin$^5$} 
\address{ 
$^1$ Dep. of Earth, Atmospheric and Planetary Sciences,
Massachusetts Institute of Technology, Cambridge, MA 02139, USA\\
$^2$ Center for Polymer Studies and Department of Physics, Boston
University, Boston, Massachusetts 02215, USA\\
$^3$ Earth and Planetary Sciences, McGill University, Montr\'eal, QC
H3A2A7, Canada\\
$^4$ Lamont-Doherty Earth Observatory of Columbia University,
Palisades, NY 10964-8000, USA\\
$^5$ Gonda-Goldschmied Center and Dept. of Physics, Bar-Ilan University, 
Ramat-Gan, Israel
}
\date{\today}
\maketitle
\begin{abstract}
  { Evidence of past climate variations are stored in ice and
    indicate glacial-interglacial cycles characterized by three
    dominant time periods of 20kyr, 40kyr, and 100kyr. We study the
    scaling properties of temperature proxy records of four ice cores
    from Antarctica and Greenland.  These series are long-range correlated
    in the time scales of 1-100kyr. We show that these series are
    nonlinear as expressed by 
    volatility correlations and a broad multifractal spectrum. We
    present a stochastic model that captures the scaling and the
    nonlinear properties observed in the data.  }
\end{abstract}
\ifnum \tipo = 2
]
\fi

\def\figureI{
\begin{figure}[thb]
\centerline{\psfig{figure=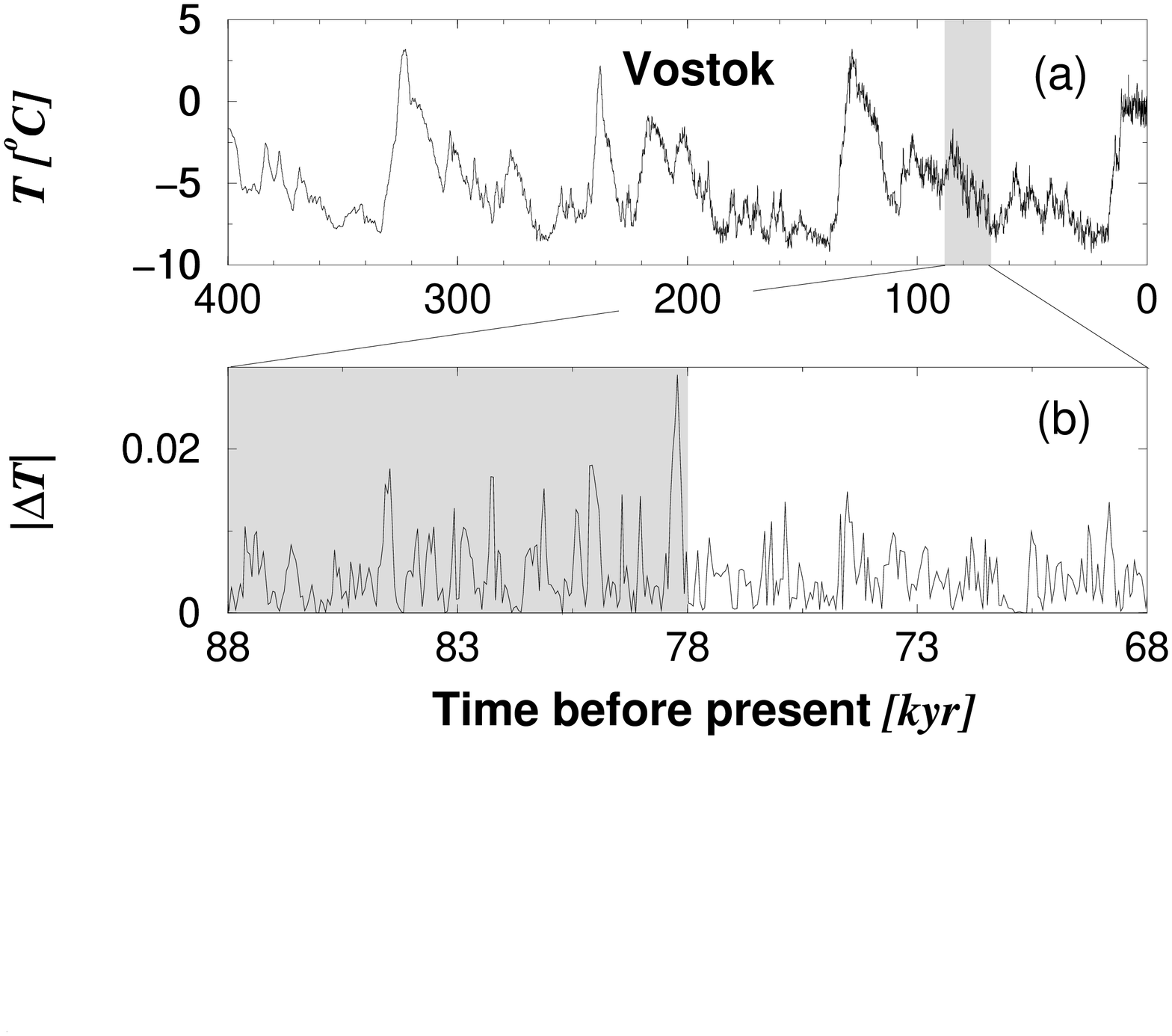,width=\figsiz1,angle=-90}}
{
\ifnum\tipo=2
\vspace*{0.0truecm}
\fi
\caption{\label{fig1}
  (a) Isotopic temperature record from the
  Vostok ice core \protect\cite{Petit}; the temperature
  $T$ is calculated 
  from the hydrogen isotope ratio.  (b) A typical example of the 
magnitude of temperature changes $|\Delta T_i|$. 
  The magnitude series is clustered --- big magnitudes are likely to be
  followed by big magnitudes suggesting the presence of
  correlations in the $|\Delta T_i|$ series.  }}
\end{figure}
}
 
\def\figureII{
\begin{figure}[thb]
\centerline{\psfig{figure=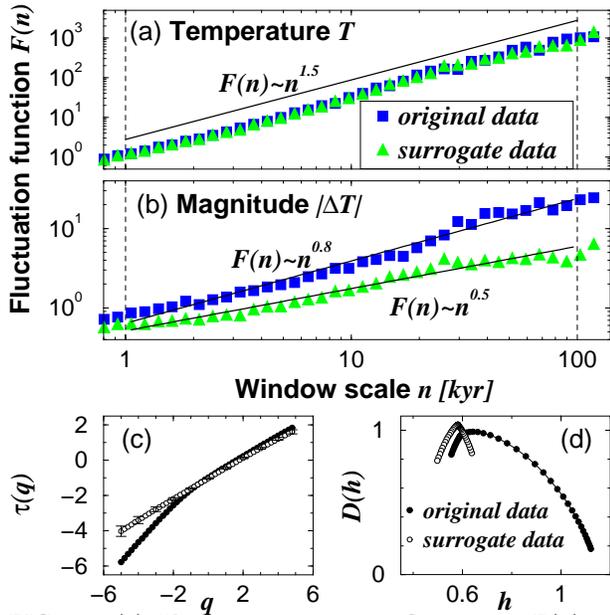,width=\figsiz1,angle=-90}}
{
\ifnum\tipo=2
\vspace*{0cm}
\fi
\caption{\label{fig2}
  (a) The root mean square fluctuation $F(n)$ as a function
  of window scale $n$ in kyr for the Vostok temperature proxy
  data indicates strong correlations (\FullBox). Before applying
  the second order DFA we sampled the data at 0.1kyr intervals
  \protect\cite{remarkDFA}. 
  The surrogate series (gray triangles)
  exhibits almost identical scaling confirming that correlations in
  the $T$ series are a linear measure.  (b) $F(n)$ for the magnitude
  series, $|\Delta T|$, indicates strong correlations (\FullBox)
  \protect\cite{Ashkenazy2001}. 
  The magnitude series of the surrogate data (gray triangles) is
  uncorrelated (with exponent 0.5) demonstrating the nonlinearity of the
  data. 
  (c) The multifractal analysis uses the wavelet
  transform modulus maxima method \protect\cite{Arneodo94}, with the
  8-tap Daubechies discrete wavelet transform
  \protect\cite{Daubechies}.  The exponents $\tau(q)$ are estimated by
  scaling function $Z_q(n) \sim n^{\tau(q)}$ \protect\cite{remarkMF}
  (0.8kyr $\le n \le $25.6kyr). 
  The curvature in $\tau(q)$ reflects the multifractality of the temperature
  series ($\bullet$). The $\tau(q)$ of the surrogate series ($\circ$)
  is linear
  (10 realizations; the
  average $\pm$ 1 standard deviation is shown).  (d) The multifractal
  spectrum, $D(h) \equiv hq-\tau (q)$ ($h \equiv d\tau /dq$), is much
  broader for the original data ($\bullet$) compared to the average
  $D(h)$ of the surrogate data ($\circ$) confirming that
  the underlying dynamics is nonlinear.  }}
\end{figure}
}
 
\def\figureIV{
\begin{figure}[thb]
\centerline{\psfig{figure=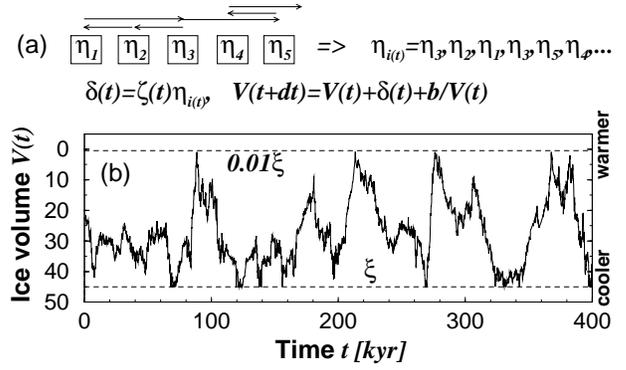,width=\figsiz1,angle=-90}}
\caption{\label{fig4}
(a) An illustration of our model. Random switching between states $\eta_i$
determines the current $\eta_{i(t)}$ that interact with $\zeta(t)$ to
change ice volume.
The ice-sheet grows ($b=1$) until it crosses a critical ice volume
$\xi$, where it breaks up ($b=-3$). Once the 
ice-sheet is totally melted ($V=0.01\xi$) growth starts again.
(b) An example of a time series generated by our model; note that the
y-axis is inverted to allow easier comparison with
Fig.~\protect\ref{fig1}. The dashed lines indicate the maximal ($\xi$)
and minimal ($0.01\xi$) ice-volume. 
}
\end{figure}
}
 
\def\figureVI{
\begin{figure}[thb]
\centerline{\psfig{figure=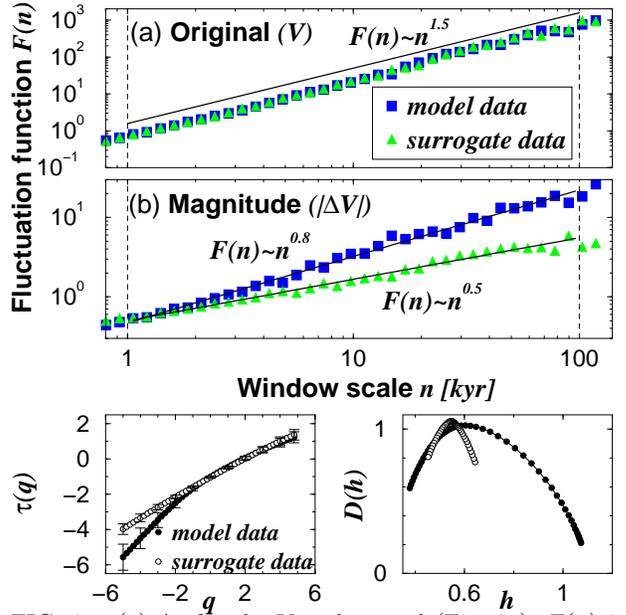,width=\figsiz1,angle=-90}}
\caption{\label{fig6}
(a) As for the Vostok record (Fig.~\protect\ref{fig2}a), $F(n)$
indicates random walk behavior (\FullBox). $F(n)$ remains unchanged
after the surrogate data test (gray triangles).  
(b) $F(n)$ for the magnitude series reproduces the correlations
observed for natural data (Fig. \protect\ref{fig2}b).  After the
surrogate data test the magnitude series becomes uncorrelated
indicating nonlinearity of our model.  
(c) The exponents $\tau (q)$ versus the moment $q$ for series
generated by the model (10 realizations of 410kyr each, the average
$\pm$ 1 standard deviation is shown) before ($\bullet$) and after
($\circ$) the surrogate data test.  The exponents measured for window
scales between 0.8kyr and 25.6kyr.  The nonlinear dependence of
$\tau(q)$ on $q$ is similar to the natural data
(Fig.~\protect\ref{fig2}c) and becomes linear after the surrogate data
test.  
(d) The multifractal spectrum $D(h)$ versus the exponent $h(q)$. As in
the data (Fig. \protect\ref{fig2}d) also here $D(h)$ is broad before
applying the surrogate data test and becomes narrower afterward.  }
\end{figure}
}

\def\TABLE{
\begin{table}
\caption{\label{table1}
Scaling results of the cores under consideration
(0.1kyr sampling {\protect\cite{remarkDFA}}). The DFA exponents
for the original ($\alpha_T$) and magnitude ($\alpha_{|\Delta T|}$)
series are obtained for window scales between 1kyr and $\sim$ 1/4 of
the series length. The multifractal analysis is not accurate for short
series, GISP, GRIP, Taylor Dome, and is
not presented; nonetheless, the multifractal spectrum for these
cores is broad as for Vostok (Fig.~\protect\ref{fig2}).
}
\begin{tabular}{ccccc}
measure & GISP & GRIP & Taylor & Vostok \\ \hline
age & 110kyr & 225kyr & 103kyr & 422kyr \\
$\alpha_{T}$ & 1.14 & 1.18 & 1.4 & 1.54 \\
$\alpha_{|\Delta T|}$ & 0.77 & 0.82 & 0.8 & 0.78 \\
\end{tabular}
\end{table}
}

Abundant geological evidence indicates that temperatures varied from
the cold of ice ages to the warmth of interglacial periods. In the
last 800,000 years (800kyr) there is strong evidence for a dominant
glacial-interglacial cycle of 100kyr, with weaker secondary cycles of
40kyr and 20kyr \cite{Petit}. Each 100kyr cycle consists of gradual
cooling for $\sim$ 90kyr followed by rapid warming during $\sim$
10kyr.  ``Milankovitch forcing'', which refers to changes in
insolation (solar radiation) due to variations in the precession,
obliquity (tilting), and eccentricity of Earth's orbit
\cite{Imbrie92} are thought to play an important role in glacial
dynamics.  These orbital variations are characterized by periods of
20kyr, 40kyr, and 100kyr, respectively.  The 20kyr and 40kyr periods
in the climate records are generally believed to be a linear response
of the climate system to insolation variations. In contrast, the
weakness of the variations in solar radiation at the 100kyr timescale
has lead to the generally accepted conclusion that the
glacial-interglacial oscillations at this timescale are most likely
not a direct linear response of the climate system to external solar
variations \cite{Imbrie92}.

Many deterministic theories have been developed to explain the
glacial-interglacial 100kyr variability; the majority suggest that the
100kyr period is a result of self-sustained nonlinear mechanisms (see,
e.g., \cite{Imbrie92,Saltzman,Gildor,Tziperman}). Other studies
proposed that climate variations are stochastic and follow scaling
laws --- the Milankovitch periods are second-order perturbations (e.g.
\cite{Kominz79,Pelletier97,Wunsch}). Importantly, both the deterministic
and stochastic mechanisms still assume that the variations on time
scales below 100kyr down to 10kyr are linear. The objectives of the
present study are to quantify the degree of nonlinearity of climate
dynamics within the time scales of 1-100kyr and to provide statistical
characteristics of the proxy records which can serve as a test for
distinguishing between existing climate models
\cite{Wunsch}.

We study the correlation (scaling) properties of climate records of
the past 420kyr.  We show that temperature variations are long-range
correlated suggesting that the Milankovitch periods are indeed
secondary and (contrary to common belief \cite{Imbrie92}) that climate
dynamics of all time scales below 100kyr down to 1kyr are highly
nonlinear.  In addition, we quantify the degree of nonlinearity in the
climate records and suggest a possible stochastic nonlinear mechanism
for our findings.

Our analysis is based on isotope records obtained from four ice cores,
Vostok and Taylor Dome from Antarctica, and GISP
(Greenland-Ice-Sheet-Project) and GRIP (Greenland-Ice-Project) from
Greenland \cite{remark1}. Measurements of oxygen and hydrogen isotope
ratios ($\delta^{18}O$ and $\delta D$) of the ice at different depths
in the core provide a proxy record of temperature
\cite{Jouzel} when the ice was formed (Fig.~\ref{fig1}a).  These
records extend back to 100-420kyr. 

Fourier analysis is the standard method for studying long-range
correlations in time series. When the power spectrum follows scaling
laws, $S(f) \sim 1/f^\beta$ ($\beta>0$), the series is long-range
correlated \cite{Shlesinger}.  However, the power spectrum might yield
an inaccurate estimation of the scaling exponent due to constant or
polynomial trends that are not necessarily related to the intrinsic
dynamics \cite{Peng94}. We therefore use the detrended fluctuation
analysis (DFA) \cite{Peng94}; the $m{\rm th}$ order DFA eliminates
polynomial trends of order $m-1$ from the data and provides a more
accurate estimation of the scaling exponent
\cite{Peng94,remarkDFA}. If the root mean square fluctuation function,
$F(n)$, is proportional to $n^\alpha$, where $n$ is the window scale,
the series is long range correlated ($\beta=2\alpha-1$).  For a random
series $\alpha=0.5$ while for correlated (or anticorrelated) series
$\alpha>0.5$ (or $\alpha<0.5$).  We begin our analysis with the Vostok
ice core and find that temperature changes are highly correlated in
the time range 1-100kyr with a scaling exponent $\alpha \approx 1.5$
(Fig.~\ref{fig2}a), consistent with the previously reported power
spectrum exponent $\beta=2$ \cite{Kominz79,Pelletier97,Wunsch}.

Next, we analyze the nonlinear properties of the ice core record. We
define a process to be {\it linear} if it is possible to reproduce its
statistical properties (such as the third moment) from the power
spectrum and the probability distribution alone, regardless of the Fourier
phases \cite{Schreiber00}.  This definition includes autoregression
processes ($x_{n}=\sum_{i=1}^Ma_ix_{n-i}+\sum_{i=0}^Lb_i\eta_{n-i}$ where
$\eta$ is Gaussian white noise) and fractional Brownian motion; the
output, $x_n$, of these processes may undergo monotonic nonlinear
transformations $s_n=s(x_n)$ and still be linear. Processes which are
not linear are defined as {\it nonlinear} \cite{remarkMF}.

Long-range correlations in the temperature time series, $T_i$, reflect
linear aspects of $T_i$. Long-range correlations in the magnitudes of
temperature increments, $|\Delta T_i|\equiv |T_{i+1}-T_i|$
(Fig.~\ref{fig1}b), which we define as volatility, indicate
nonlinearity of the underlying process
\cite{remarkMF,Ashkenazy2001}. Linear series have uncorrelated
$|\Delta T_i|$ series while nonlinear processes that follow a scaling
law exhibit long-range correlations in the magnitude series $|\Delta
T_i|$. We find that the magnitude series $|\Delta T_i|$ is highly
long-range correlated within the time range 1-100kyr
(Fig.~\ref{fig2}b). Thus, the underlying
process is nonlinear \cite{Ashkenazy2001}. The value of the
correlation exponent quantifies the degree of nonlinearity in the ice
core record. Correlations in the magnitude series indicates that the
magnitude series is ``clustered'', i.e., large magnitude is more
likely to be followed by a large magnitude, as can be seen in
Fig.~\ref{fig1}b. These clusters may be associated with abrupt warming
events known as Dansgaard-Oeschger events \cite{Dansgaard}. 

To demonstrate that the correlations in the magnitude series are
related to the nonlinearity of the underlying process we apply a
surrogate data test for nonlinearity that preserves both the power
spectrum and the histogram of the temperature increment series $\Delta
T_i$ \cite{Schreiber00}.  The surrogate series has random Fourier
phases; the nonlinearities that are stored in the phases are
destroyed. We find that the magnitude series obtained from the
surrogate series is indeed uncorrelated (Fig. \ref{fig2}b) confirming
that the original series is nonlinear within 1-100kyr.

Correlations in the magnitude series $|\Delta T_i|$ can be related to
the width of the multifractal spectrum
\cite{remarkMF,Ashkenazy2001}.
We calculate the exponents $\tau(q)$ of different moments $q$ for the
ice core data and find that $\tau(q)$ is a nonlinear function of $q$
(Fig. \ref{fig2}c), indicating that the temperature series is
multifractal. We also perform multifractal analysis on the surrogate
data and find that its $\tau(q)$ is almost linear.  The multifractal
spectrum, $D(h)$ (Fig.~\ref{fig2}d), is broad for the original data
and narrower for the surrogate data. The broadness of the multifractal
spectrum may also be used to quantify the degree of multifractality,
and thus the degree of nonlinearity, in the data \cite{Ashkenazy2001}.

We repeat the above analysis for the other three ice cores and obtain
similar results (Table \ref{table1}).  Although the DFA exponents of
the original series are smaller for the Greenland cores, the magnitude
series exponents are almost the same for all cores. We thus conclude
that climate dynamics is nonlinear for time scales of few thousands of
years up to 100kyr.

To understand the mechanism that may contribute to the nonlinearities
observed in the data we modified a model for ice-volume evolution
recently suggested by Wunsch \cite{Wunsch}. Ice volume $V$ was
observed to be negatively correlated with temperature $T\sim -V$
\cite{Petit} and thus a model for ice accumulation may serve,
indirectly, as a model for temperature dynamics.  The Wunsch model can
be summarized as follows: The ice-sheet builds randomly up to a
critical volume where it breaks up rapidly.  Then, growth begins
again.  By construction, this model is a random walk up to a time
scale corresponding to the critical ice volume, followed by a
crossover to random behavior for larger time scales.  However, this
model does not reproduce the nonlinearity in the data as defined and
found above.

The assumptions of our model (Fig.~\ref{fig4}) are:

(i) The ice volume $V$ changes with steps $\delta +b/V$; i.e.,
$V(t+dt)=V(t) + \delta (t)+b/V(t)$.

(ii) When ice volume $V$ ``crosses'' a critical volume $\xi$, $b$ is
set to be negative, $b=b_2<0$.  
Ice volume is considered to lie between $0.01\xi$ and $\xi$. 
When $V=0.01\xi$, $b$ becomes $b=b_1>0$ till $V$ exceeds the threshold $\xi$.

(iii) The ice accumulation increments $\delta$ are the product of two 
stochastic inputs, $\delta(t)=\zeta(t) \eta_{i(t)}$. $\zeta$ and 
$\eta$ are Gaussian distributed random variables with zero mean and unit
variance. 

(iv) Random switching between the states $\eta_i$'s is controlled by $i(t)$
which is equal to $[l(t)]$ where $[\cdot ]$ stands for the closest 
integer value. $l$ is a random walk described by, $l(t+dt) = l(t) +
C\omega (t)$, where $C$ is the switching range and $\omega$ 
is another Gaussian random variable with zero mean and unit variance
(see Fig.~\ref{fig4}a).

Assumptions (i) and (ii) describe the random growth (with $b=b_1>0$)
of the ice-sheet and its rapid breakup (with $b=b_2<0$)
after crossing the critical volume $\xi$. The $1/V$ term in assumption
(i) mimics the reduced ice accumulation for large ice-volume
\cite{Gildor}. Ice volume changes $\delta$ result from two interacting
random inputs [assumption (iii)] 
where one, $\zeta$, may represent the atmosphere, the other, $\eta$, the
ocean, and the product, $\zeta\eta$, the atmosphere-ocean interaction. 

In our simulations of the model we use the following values,
$dt=0.1kyr$, $\xi=1.5 \sqrt{90kyr/dt}$, $b_1=1$, $b_2=-3$, and
$C=0.27$. We choose the value 90kyr in $\xi$ so that on average $V$
will grow from zero to $\xi$ after 90kyr/dt steps.  
We choose the values of $b_1$ and $b_2$ such that the ice-sheet grows
slowly and breakup rapidly.
The values of $\xi$, $b_1$, and $b_2$ are constrained by the natural
record. The switching parameter $C$ determines the number of states
$\eta$ for a given number of steps; the number of different states is
proportional to the square root of the number of steps (e.g., $\sim3$
states for 4kyr and $\sim5$ for 10kyr). 

An example of an arbitrary 400kyr time series obtained by the model is shown in
Fig.~\ref{fig4}b. The scaling of the model's $V$ series
(Fig.~\ref{fig6}a) indicates random walk behavior with exponent
$\alpha=1.5$ (as for the Vostok core).  The magnitude series $|\Delta
V|$ is highly correlated with exponent $\sim 0.8$ (Fig.
\ref{fig6}b) as for the ice core data (Fig.~\ref{fig2}b and
Table~\ref{table1}). The surrogate data test applied to the $\Delta V$
series changes the magnitude series into an uncorrelated one,
indicating the nonlinearity of the model.  This nonlinearity is mainly
due to the product of the inputs $\eta$, $\zeta$ in assumption
(iii). The multifractal spectrum is broad where, as with the ice core
data (Fig. \ref{fig2}c,d), the exponents for negative moments, $\tau
(q<0)$, mainly contribute to its broadness (Fig. \ref{fig6}c,d). After
the surrogate data test the series becomes linear and statistically
different from the original data.

This simple model reproduces the statistical characteristics of the
ice core data under consideration. Although the natural system is
undoubtedly more complex, we conjecture that the model variables may
be associated with specific aspects of Earth's climate system although
our model cannot uniquely identify them.  One of the random inputs,
$\eta$, thus may represent the influence of the deep  
ocean on ice accumulation since the state of the deep ocean is known
to have impact on glaciation (e.g., \cite{Tziperman}). The other
random input, $\zeta$, may represent the net atmospheric influence
affecting ice accumulation (resulting from, e.g., variations in eddy
transport, cloudiness, incoming solar radiation, ablation).  We assume
that the ocean has several states with a tendency to return to
previous states, as does $\eta$ in the model; a possible example for
such ``switching'' mechanism is the deep ocean circulation which has
few states with possible switching between them (e.g. \cite{REFS}).

We conclude that climate changes in the time range of 1-100kyr are
long-range correlated confirming the major role of stochasticity in
climate \cite{Kominz79,Pelletier97,Wunsch}. Moreover, our results
suggest that the underlying dynamics in the time scales of 1-100kyr is
nonlinear. This nonlinearity is specified and quantified by strong
long-range correlations in the magnitudes of temperature changes and
in a broad multifractal spectrum. Our simple stochastic model suggests
that the nonlinearity can be the result of only two random processes
that interact with each other. Glaciation models may be generally
categorized into two main alternatives: (i) linear mechanisms that are
driven by stochastic forcing (e.g. \cite{Pelletier97,Wunsch}), and
(ii) nonlinear mechanisms without stochastic forcing
(e.g. \cite{Saltzman,Gildor}). Our results and model suggests a third
alternative --- nonlinear mechanism that inherently involves
stochastic forcing.  Our results raise a new challenge for the many
climate models, and may help guide development of better climate
models, which include both periodic and stochastic elements of climate
change.

YA and HG thank the Bikura fellowship for financial
support. DRB thanks H.E. Stanley for his
generous hospitality. We thank P. Cerlini, P. Hybers, V. Schulte-Frohlinde, P.H. Stone, E. Tziperman, 
and
C. Wunsch for
helpful discussions.

\ifnum\tipo=1
  \figureI
  \figureII
  \TABLE
  \figureIV
  \figureVI
\fi
\ifnum\tipo=2
  \figureI
  \figureII
  \TABLE
  \figureIV
  \figureVI
\fi


\begin{references}
\bibitem{remarkDFA} Before applying the scaling technique we evenly sampled
the ice core data (50yr and 100yr). We find very similar DFA exponents
for the different sampling rates. 
The scaling regime starts at 1kyr and is larger than the maximal 
spacing between consecutive values.
\end{references}
\end{document}